# Independent evaluation of the significance of the recent ATLAS and CMS data

Gioacchino Ranucci

*Istituto Nazionale di Fisica Nucleare*
*Via Celoria 16 - 20133 Milano*
*Italy*
*Phone: +39-02-50317362*
*Fax: +39-02-50317617*
*e-mail:* gioacchino.ranucci@mi.infn.it

**Abstract**

This note describes an independent assessment of the statistical significance of the recently released ATLAS and CMS data, about 11 $fb^{-1}$ per experiment acquired in 2011 and in the first part of 2012, for what concerns the Higgs search in the two high resolution decay channels especially suited for the low mass region, i.e. the diphoton and four-lepton decay channels. Scope of this note is not to reproduce the analysis of the Collaborations: this would be impossible given the enormous complexity of the complete profile likelihood procedure used to evaluate local and global the p-values, and the huge number of nuisance parameters which are used to incorporate the numerous systematic effects. Rather, its purpose is to show the significance that an outsider can infer only on the basis of the released data and plots, used as input of a simplified profile likelihood procedure in which the only contemplated nuisance parameter is the background normalization in the diphoton channel. In practice, this note tries to address the question of the independent judgment of the significance of new data that physicists are used to perform on their own when they are shown for the first time particularly relevant results with indication of new effects, and that in the complex LHC framework is not so easily doable as in other experimental contexts.

*Keywords:* Profile likelihood ratio; P-value; Local and Global Significance; Look elsewhere effect; Higgs search

*PACS:* 06.20.Dk

## 1. Introduction

The recent release of the LHC data has been characterized by an interesting "psychological" effect: the audience during both 4th July seminars did not applaud when the hints of interesting signals were showed by the speakers in the gamma-gamma and four-lepton distributions, but instead when the magic words "five sigma" were pronounced by the speakers themselves.

This fact shows how pervasive and persuading has been the diffusion of the 5 sigma "mantra" within the particle physics community over at least the past decade. But, if people not directly involved in the research are convinced of the validity of the result by the 5 sigma revelation, is there for these people the possibility to gauge by themselves the achievement of the 5 sigma significance level on the basis of the published data? At a first sight, this seems not possible given the huge complexity of the profile likelihood method used by the Collaborations to analyze the data, especially because of the large number of nuisance parameters which are involved in the analysis.

This is, by far and so far, an unusual situation in experimental physics, since usually when an experimental group presents its results, the audience constituted by other experimental physicists is in condition to judge autonomously and independently the significance of the published data.

Scope of this note is to address the previous question mark by performing an independent exercise to understand the significance which an outsider can compute only on the basis of the data disclosed at the CERN seminars (and in the accompanying notes).

Since the significance of the Higgs detection is stemming essentially from the gamma-gamma and four-lepton channels, the other decay modes are ignored in this exercise.

## 2. CMS

### *2.1 Gamma – gamma*

In the 4th July seminar CMS showed the categories in which the gamma-gamma data have been split both in the 7 TeV (five) and 8 TeV (six) data [1][2]. The exercise in the present note is based, however, only on the inclusive distribution obtained by summing all categories. This is by far a radical approach, justified in the spirit of understanding what an outsider can do without incurring in too hard complications, like the joint analysis of the categories taken separately, which is instead the standard of the Collaboration analysis.

Fig. 1 reports the inclusive gamma-gamma distribution obtained by summing all the categories presented by the Collaboration, while Fig. 2 displays the distribution that the Collaboration has obtained with a weighted sum (where the weights depend upon the S/B ratio) of the categories, in order to enhance visually the irregularity around 125-126 GeV, visible also in Fig. 1.

The determination of the significance of the bump which can be appreciated in both plots requires the appropriate modeling of the background. Following the indication of the Collaboration, this is done by attempting to reproduce the background components displayed in the released plots with polynomials of order 4 or 5.

In particular, it comes out that the background component of the inclusive distribution as shown in Fig. 1 is better described by a 5 order polynomial, while that in Fig. 2 by a polynomial of order 4.

The signal, instead, is simply modeled with a Gaussian whose width has been inferred from the resolutions listed in table 2 of ref. [1]. The assumption of a Gaussian shape is a rather radical approximation, consistent however with the idea of a simplified calculation inspiring this note.

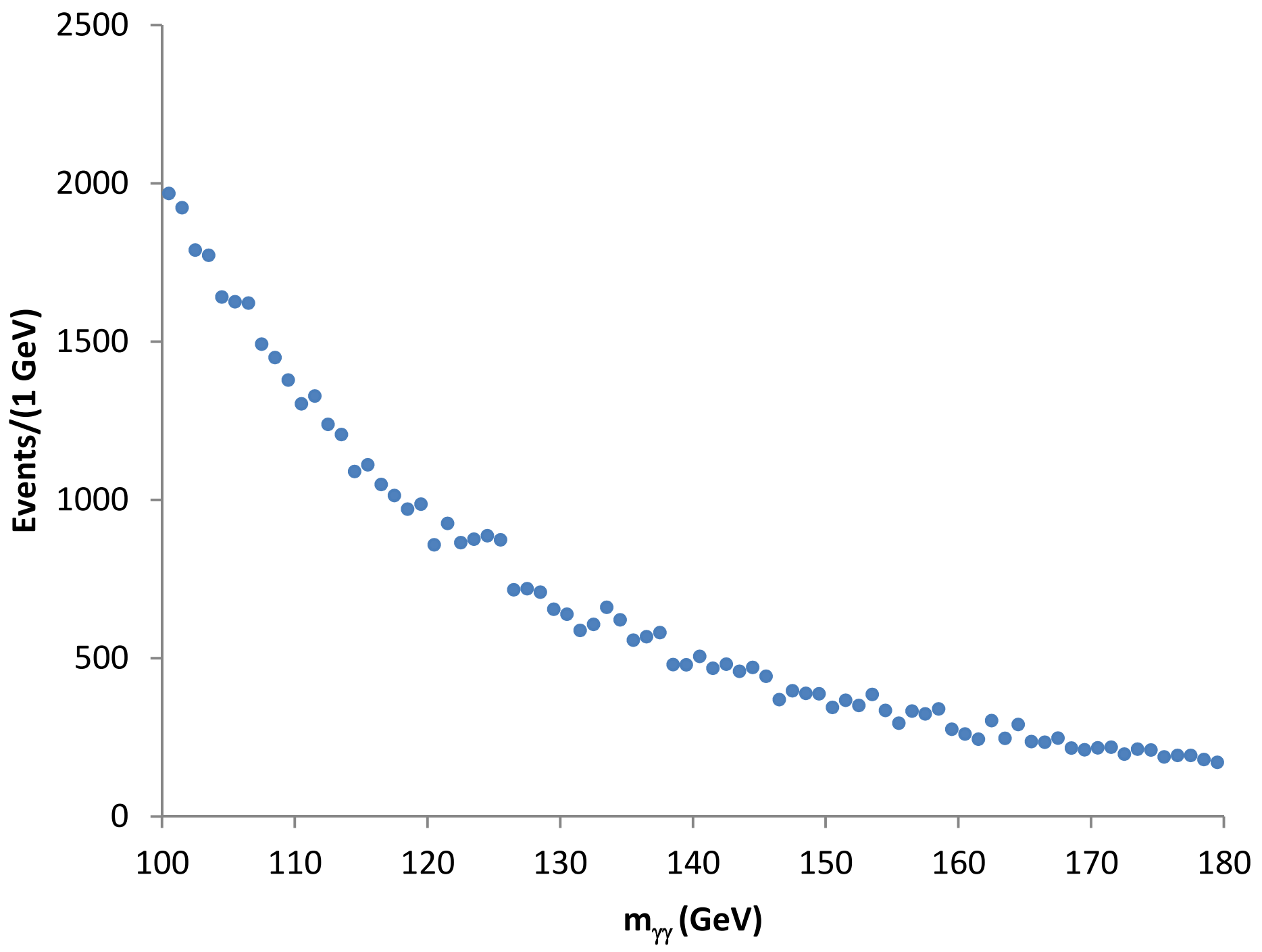

*Fig. 1 – CMS inclusive diphoton distribution*

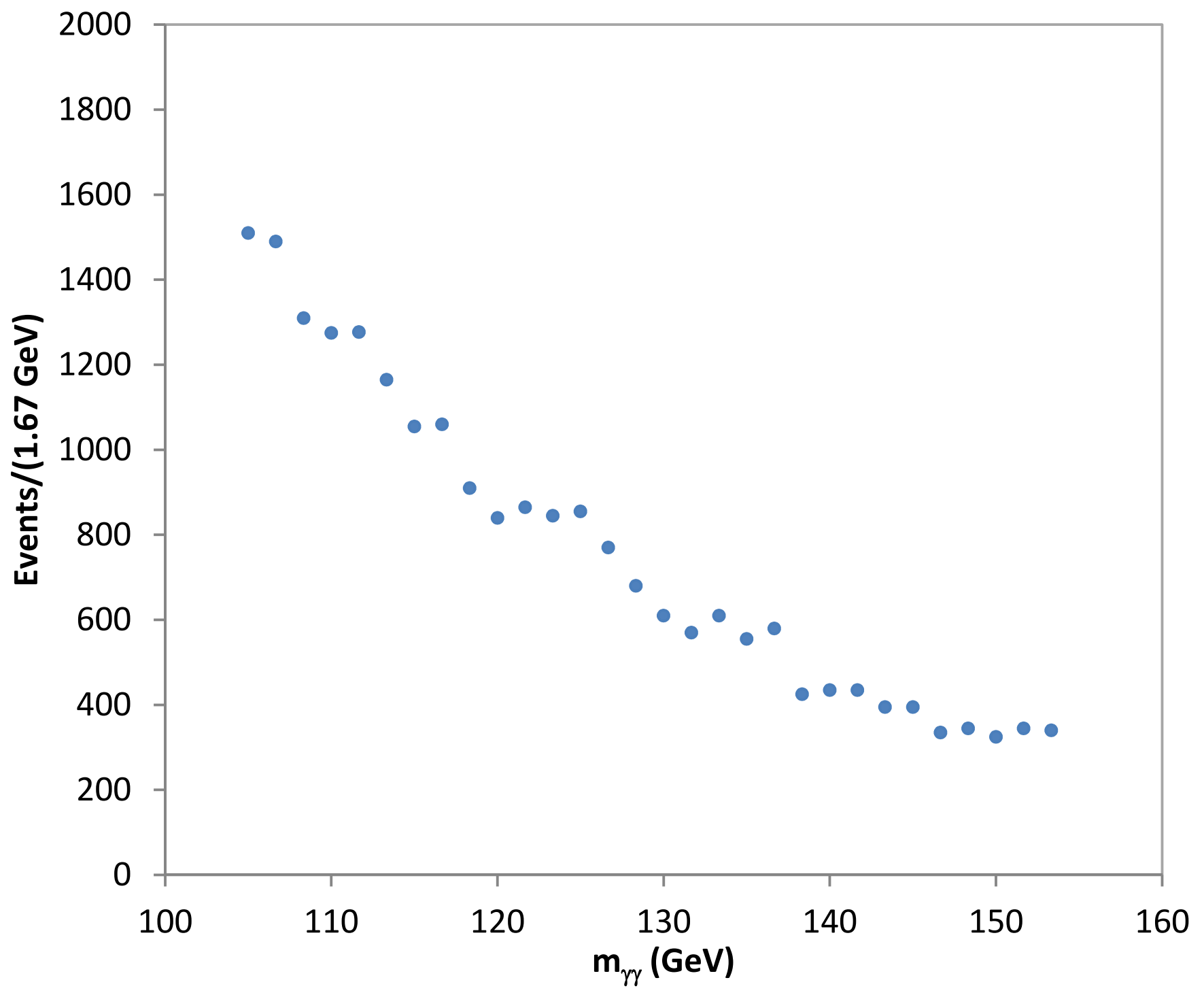

*Fig. 2 - Same CMS diphoton mass distribution as in Fig. 1, but obtained by the S/B weighted sum of the individual categories*

With these models of signal and background, the profile likelihood procedure is then accomplished for the purpose to determine the local p-value, by assuming the background normalization as nuisance parameter. First the original distribution in Fig.1 is considered. The result of the calculation performed over the low mass range from 110 to 146 GeV is shown in Fig. 3, as the yellow curve directly overlapped, for an immediate comparison, with the results presented by the Collaboration.

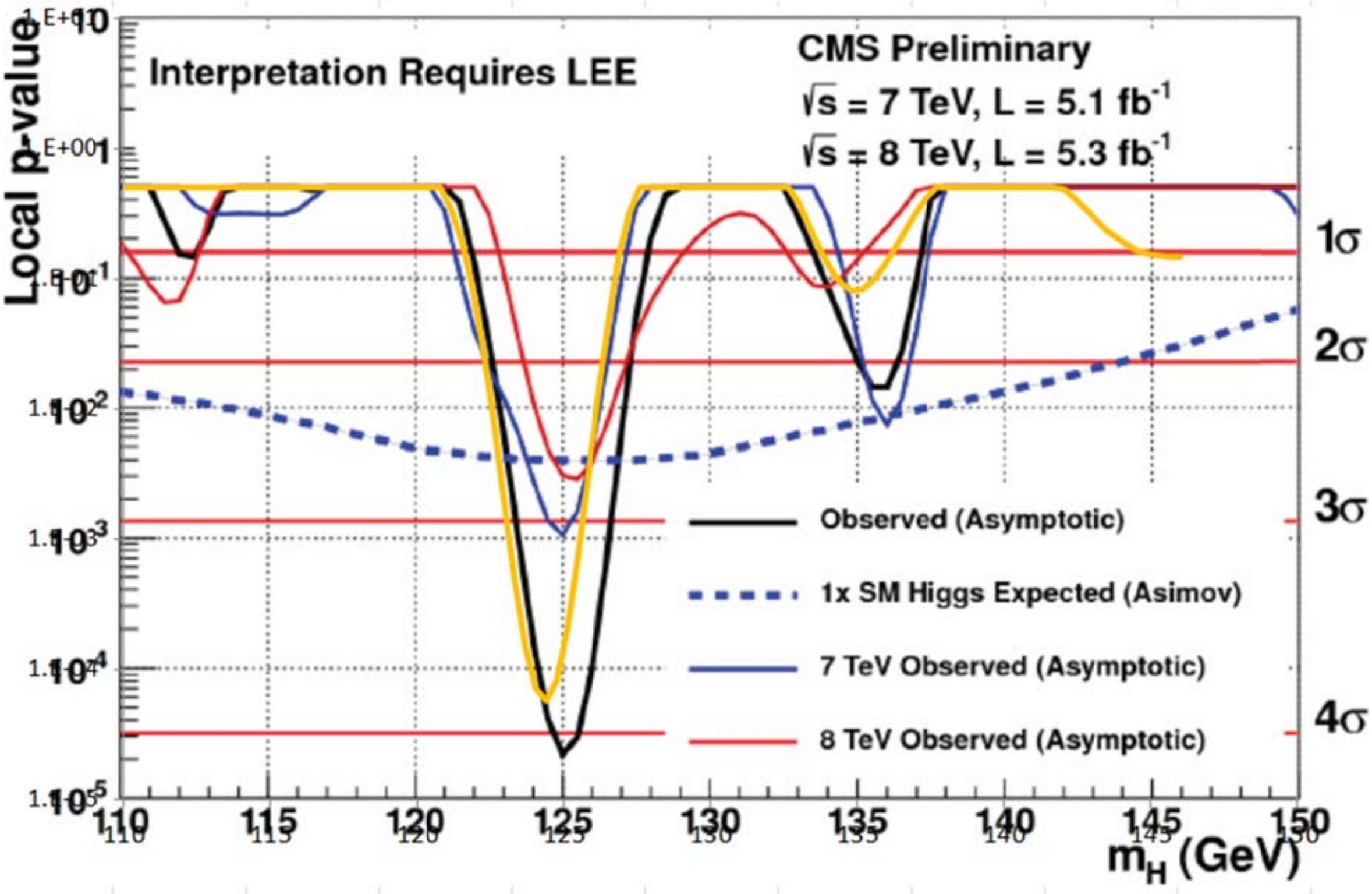

*Fig. 3 – Local p-value of the original inclusive diphoton distribution from the present evaluation (yellow line) compared with the official CMS result (black line)*

In particular the comparison is to be done with the black curve in the Collaboration plot, which is the p-value for the entire 2011+2012 data set. There is a less than 1 GeV shift in the minimum, likely to be attributed to some binning uncertainty in the data as retrieved for this note, but the p-value minima are in acceptable agreement. The published value corresponds to 4.1 σ, the value obtained with the present, simplified approach amounts to 3.85 σ.

A similar exercise, done for the weighted sum distribution in Fig. 2, is reported in Fig. 4. Still, there is a shift of 1 GeV in the minimum value, but in opposite direction, again attributable to some binning uncertainty, but what should be noted is that the minimum of the p-value is practically coincident with that reported by the Collaboration.

It can, thus, be concluded that for the gamma-gamma distribution released by the CMS Collaboration the significance that an outsider can infer from the published data is well in agreement with that obtained by the Collaboration, which by the way has been inferred not on the inclusive distribution, but through the joint analysis of the separate categories.

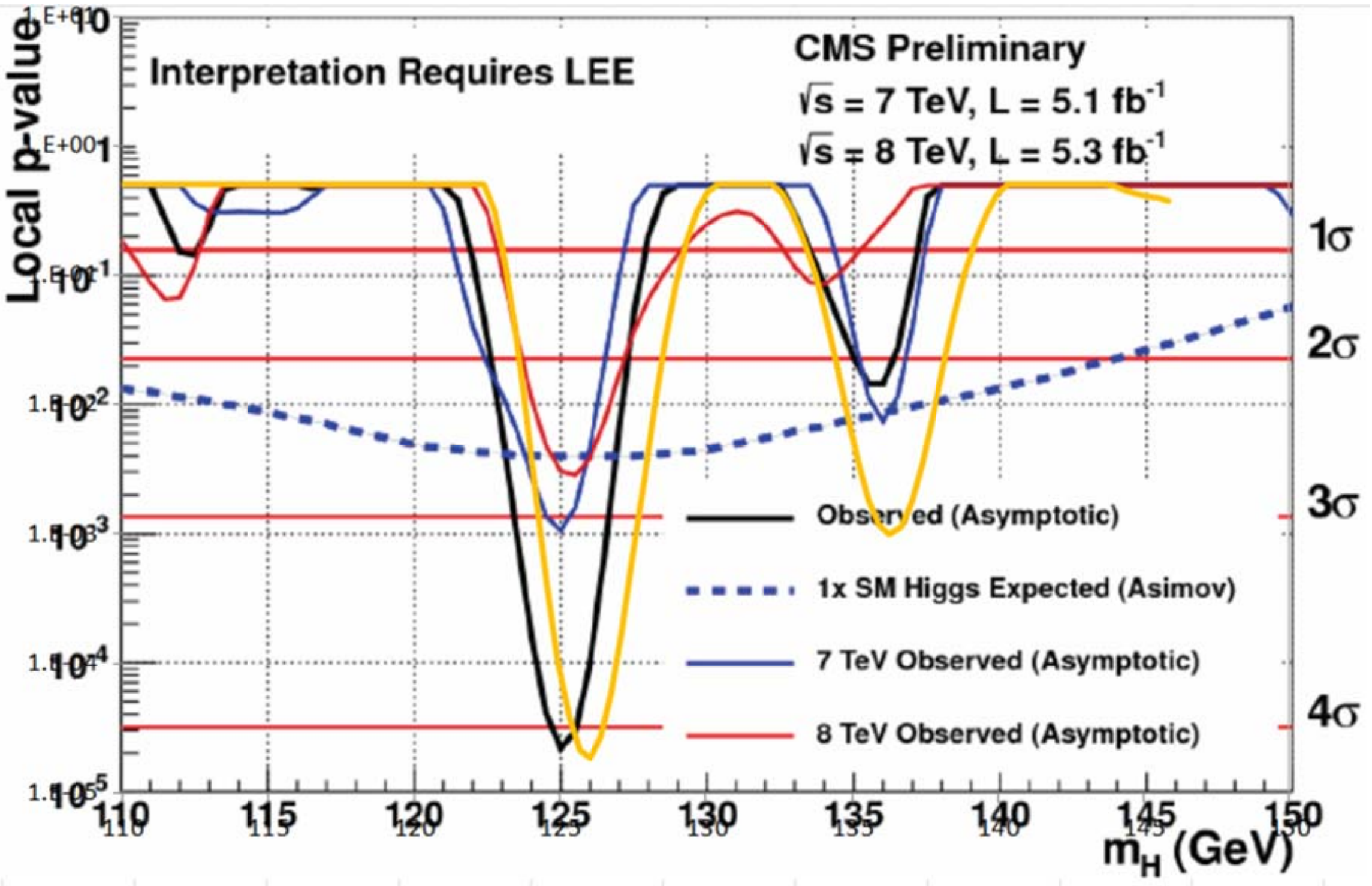

*Fig. 4 - Local p-value from the present evaluation (yellow line) of the weighted sum inclusive diphoton distribution, compared with the official CMS result (black line)*

*2.2 Four-lepton channel*

The low count statistic four-lepton distribution is shown in Fig. 5, zoomed in the low mass region of interest [3].

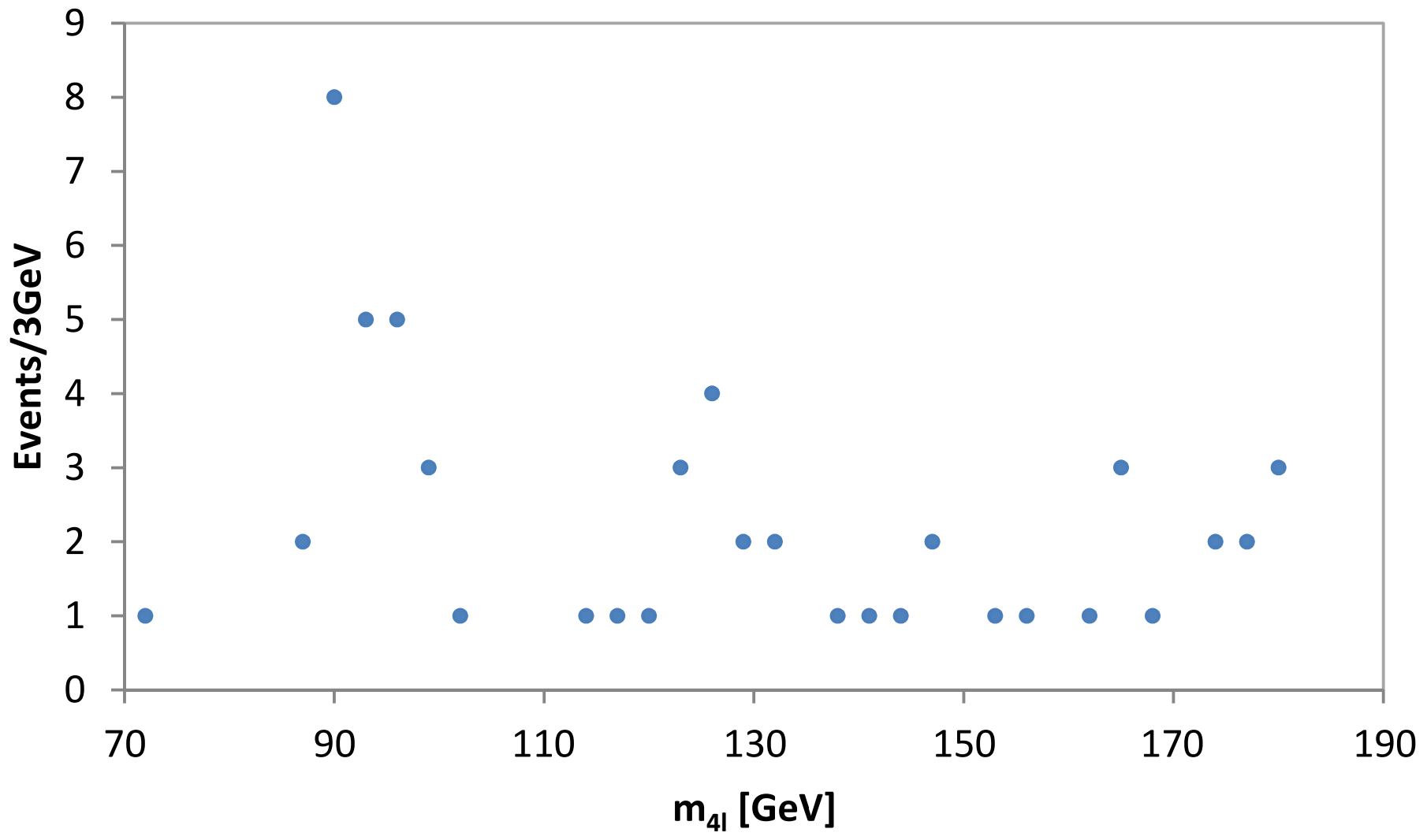

*Fig. 5 - CMS inclusive four lepton distribution*

The structure at around 90 GeV is the peak of the Z → 4ℓ "standard candle" , which is observed as expected around $m_{4l} = m_Z$ .

For the application of the profile likelihood procedure the background shape and normalization is taken from the plots in [3], while the signal is modeled by a Gaussian of width computed on the basis of the resolution information contained in [4].

In this case the straightforward implementation of the method originates a quite different result for the p-value with respect to that of the Collaboration; this can be appreciated from Fig. 6 (as before the yellow line is the output of the present calculation).

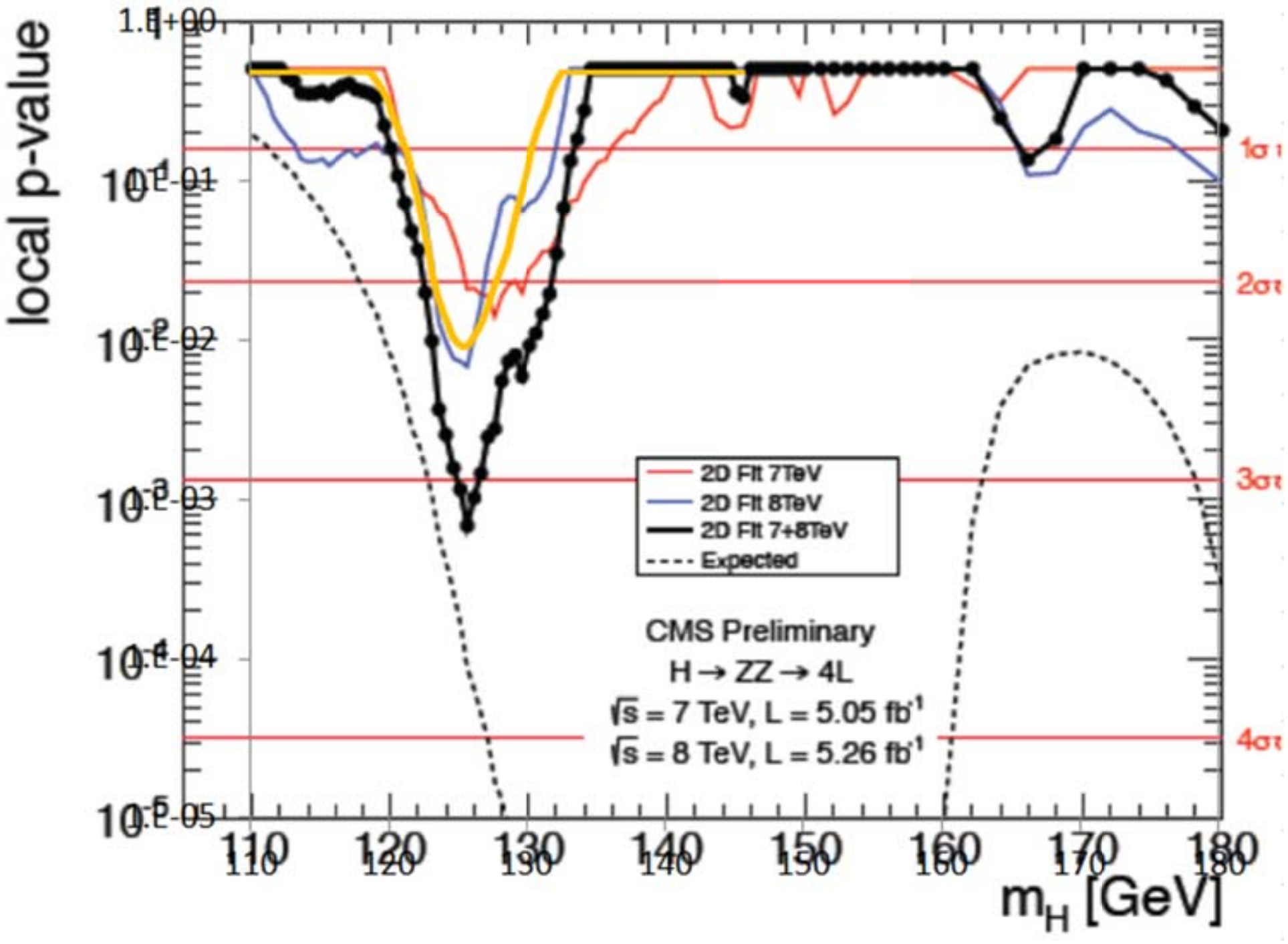

*Fig. 6 – Local p-value of the four-lepton distribution from the present evaluation (yellow line) compared with the official CMS result (black line)*

While the abscissa of the minimum of the p-value is still found in the neighborhood of 125 GeV, the significance corresponding to the minimum itself is 2.3 σ, against the 3.2 σ value (black curve) obtained by the Collaboration. This discrepancy, however, is perfectly understandable: in the note [3] it is explained that the analysis is actually performed dividing the data in three categories (according to the final state) and considering also a discrimination parameter stemming from the so called MELA background analysis; the incorporation of such a parameter makes, thus, the procedure effectively a more powerful 2D analysis.

In the same note it is reported that a 1D analysis, without accounting for the MELA parameter, gives a significance of 2.2 σ, hence in perfect agreement with the result found here. By the way, the simple Poisson counting over background of the inclusive distribution provides (obviously) a very similar significance, i.e. 2.1 σ.

*2.3 Joint local significance*

The joint local significance is computed through the concurrent application of the profile likelihood procedure to the four-lepton and gamma-gamma distributions. The result in term of local p-value is shown in Fig. 7, as usual overlapping the yellow line from this exercise to the official CMS plot (however now the CMS plot to consider for the comparison is the blue one).

Even considering the binning effect, the maximum is found at an abscissa close to 125 GeV in reasonable agreement with the official published value; the local significance is 4.6 σ, so not bad if compared with the 5 σ obtained by the Collaboration. The main contribution to this difference is likely coming from the different treatment of the four-lepton channel, that here is processed as a mere, unique counting channel.

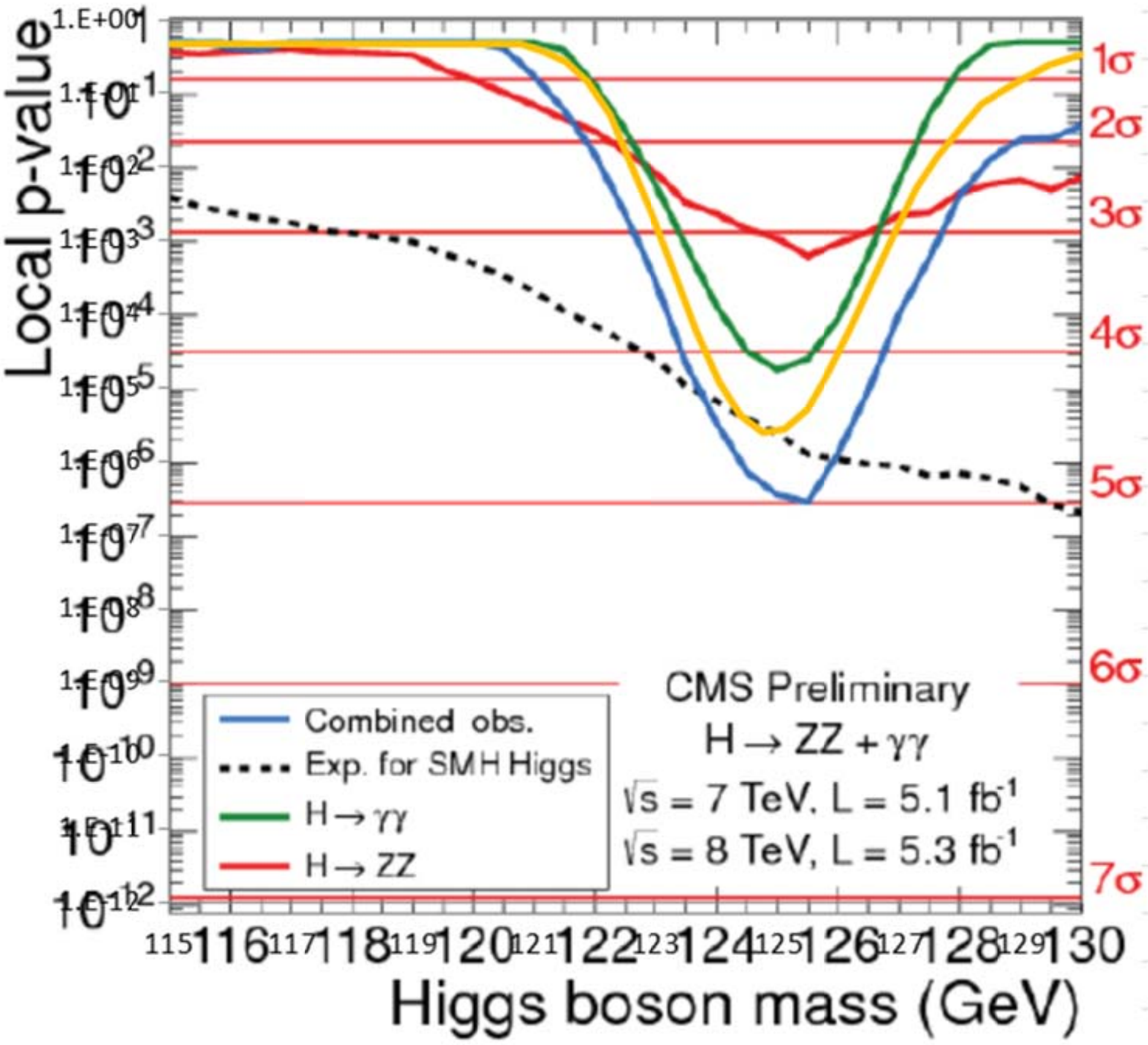

Fig. 7 - *Local p-value from the joint analysis of the four-lepton and gamma-gamma distributions through the present evaluation (yellow line) compared with the official CMS result (blue line)*

So, the simple calculation shown here demonstrates that, on the basis of the data published by the Collaboration and of a simplified implementation of the profile likelihood method, the significance that is computed is not much lower than that obtained by the Collaboration through the full calculation.

*2.4 LEE*

Given the last year exclusion limits obtained with the 2011 5 $fb^{-1}$ of data, the “prejudice” while investigating the full dataset of about 11 $fb^{-1}$ is to look for something in the neighborhoods of 125 GeV, with therefore less importance attributed to the de-rating of the significance associated with the Look Elsewhere Effect. While this attitude is practically the right one, here for the only purpose of completeness of the mathematical treatment, the Look Elsewhere Effect over the so called low mass interval of 110-146 GeV is also accomplished.

The calculation is not described in detail, since it is based on the procedure already illustrated at length in [5] and applied in the previous note [6].

The result is that in the low mass range 110-146 GeV the global significance of the joint excess in the two channels is 3.9 σ: this is thus the value to which the local significance of 4.6 σ is de-rated by the Look Elsewhere Effect. In the same mass range the Collaboration computes the significance after the LEE equal to 4.4 σ [7].

## 3. ATLAS

### *3.1 Gamma – gamma*

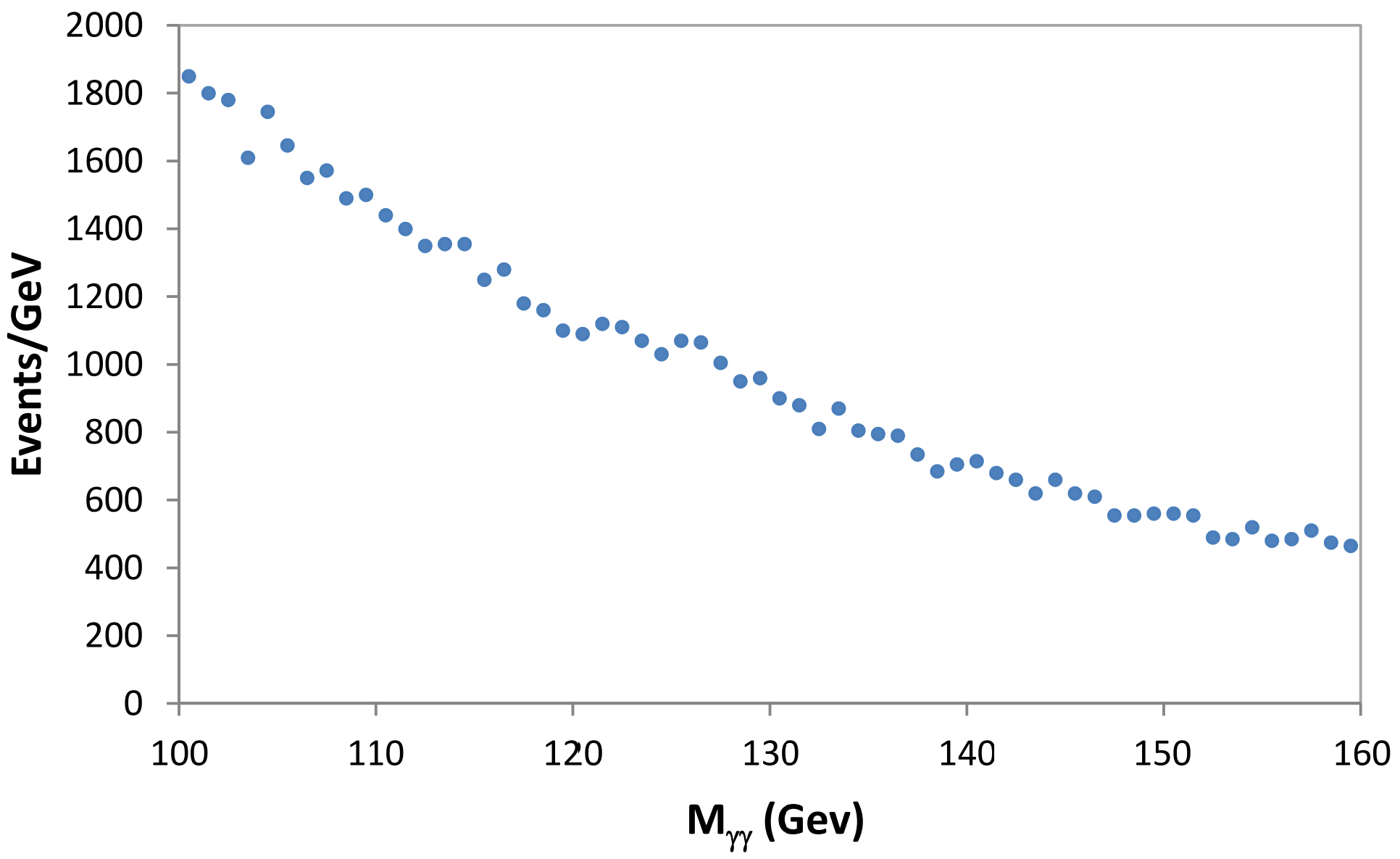

*Fig. 8 - ATLAS inclusive diphoton distribution*

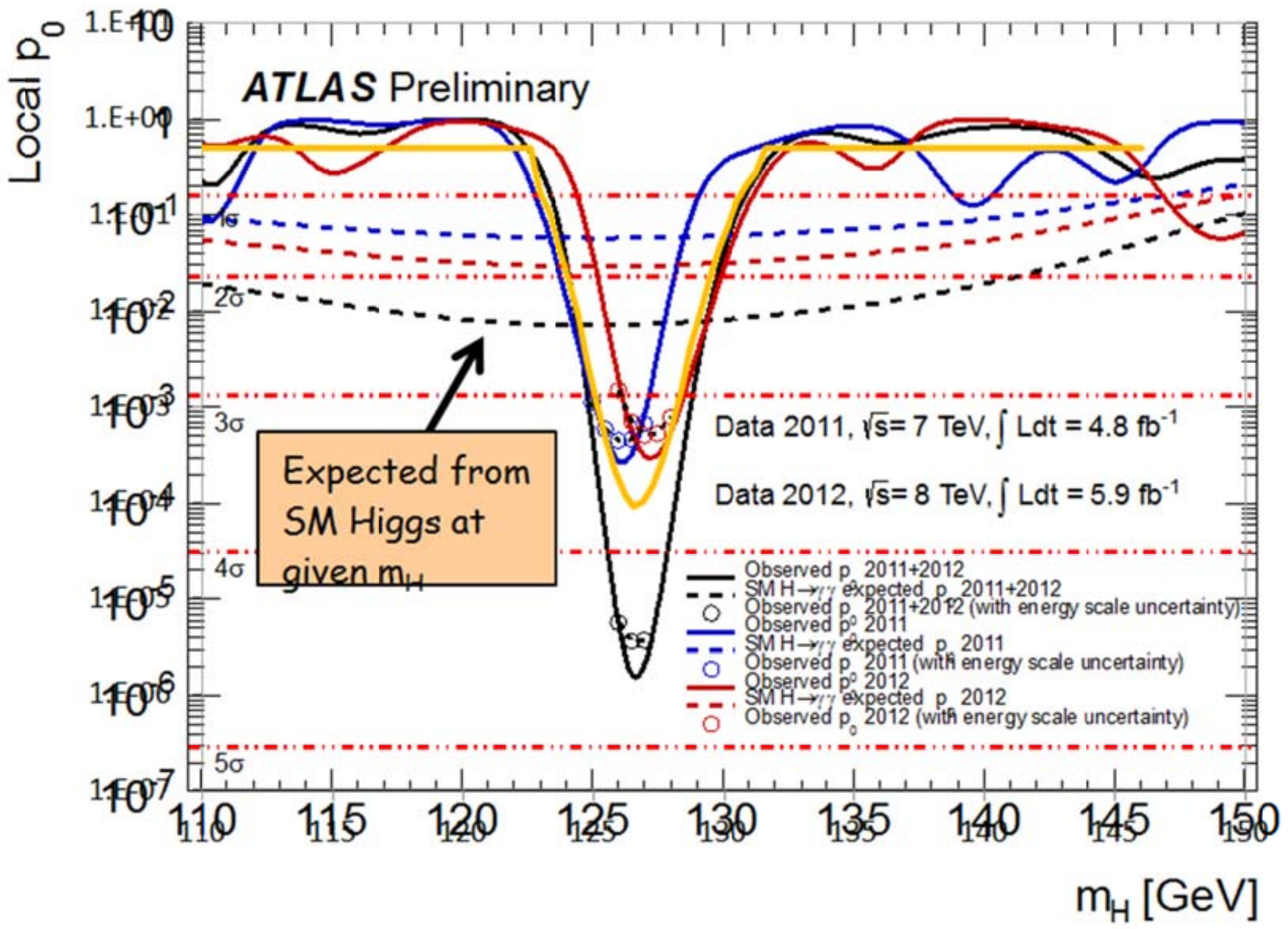

*Fig. 9 - Local p-value of the diphoton distribution from the present evaluation (yellow line) compared with the official ATLAS result (black line)*

Contrary to CMS, ATLAS at the 4th July seminar [8] did not show separately the categories of the gamma-gamma data, but only the inclusive distribution, which is reported in Fig. 8 (actually the separated categories are available in [9]).

For the background modeling in this case the Collaboration has adopted several different functions [9], depending on the category. For uniformity with the calculations performed in [6], a simple exponential model for the inclusive distribution is adopted for the present evaluation, again assuming its normalization as a nuisance parameter. Similarly to the CMS calculation, the signal is modeled as a Gaussian function with resolution taken from table 2 of ref. [9].

The p-value from the calculation performed over the low mass range from 110 to 146 GeV is shown in Fig. 9, following the usual convention of a yellow curve overlapped graphically to the official plot of the Collaboration (the reference ATLAS curve is the black one).

While the abscissas of the minima are better aligned with respect to the CMS case, the discrepancy of the minimum p-value between the official result and this exercise is decisively more pronounced: 4.6 $\sigma$ against 3.7 $\sigma$. Since the result from the Collaboration is inferred through the joint analysis of the separate categories, by comparison with the CMS case it stems that the significance of the ATLAS gamma-gamma excess is substantially boosted by the categories analysis, while this does not happen to the same degree for CMS. For comparison it can be noted that in [9] the significance obtained by the Collaboration without dividing the datasets into categories amount to 3.5 $\sigma$, very well in agreement with the 3.7 $\sigma$ value inferred here.

*3.2 Four-lepton channel*

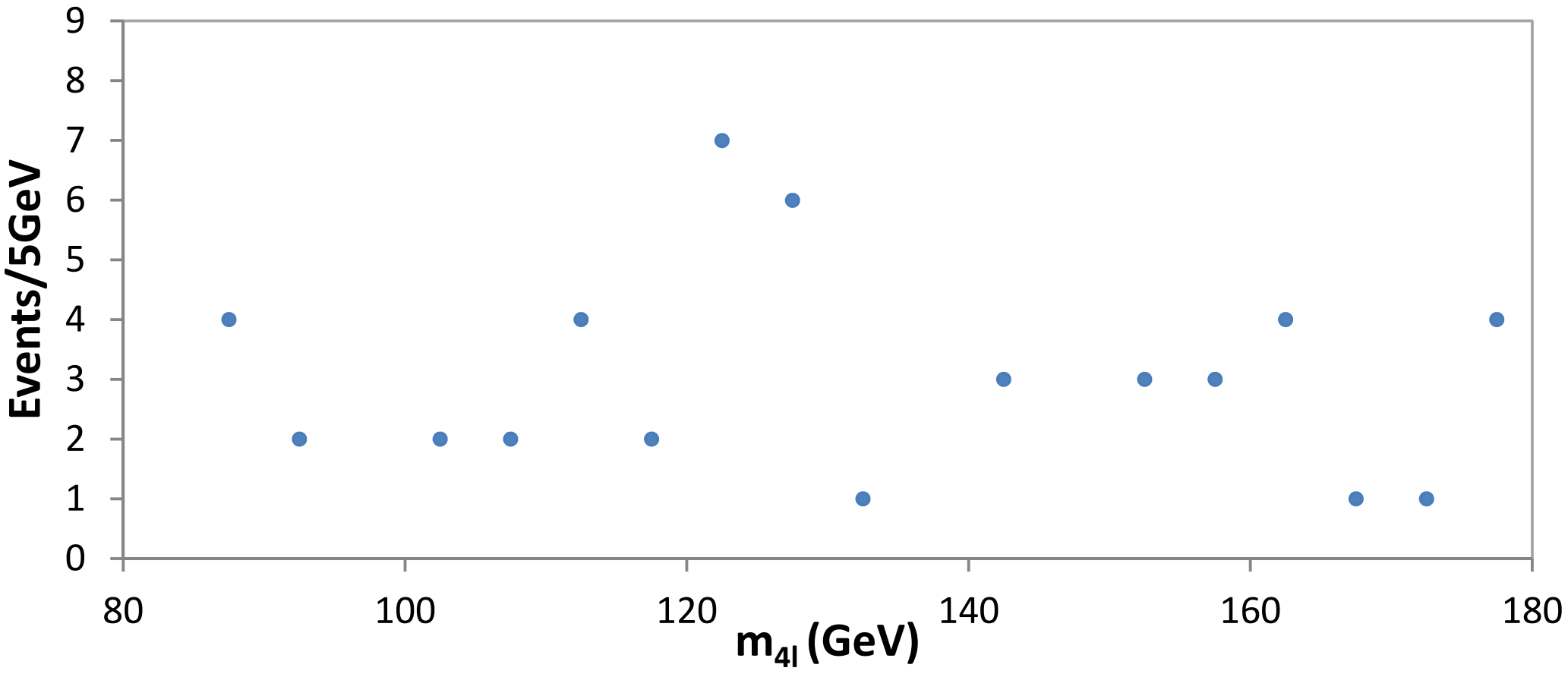

*Fig. 10 - ATLAS inclusive four-lepton distribution*

The low count statistics four lepton distribution is shown in Fig. 10, zoomed in the low mass region of interest. The background information, shape and normalization, needed for the profile likelihood calculation is taken from the plots in [10], as well as the width of the Gaussian signal model.

The resulting p-value (yellow curve in Fig. 11) features a minimum around 125 GeV corresponding to a significance of 2.9 $\sigma$, in agreement with the simple Poisson counting evaluation, that would lead to 2.8 $\sigma$. The stronger significance of 3.4 $\sigma$ obtained by the Collaboration is due, as in the previous diphoton case, to the more powerful categories analysis.

In passing, it can be noted that the boost of the significance obtained by ATLAS through the categories analysis is less powerful than that got by CMS; likely this occurrence is linked with the CMS MELA background analysis.

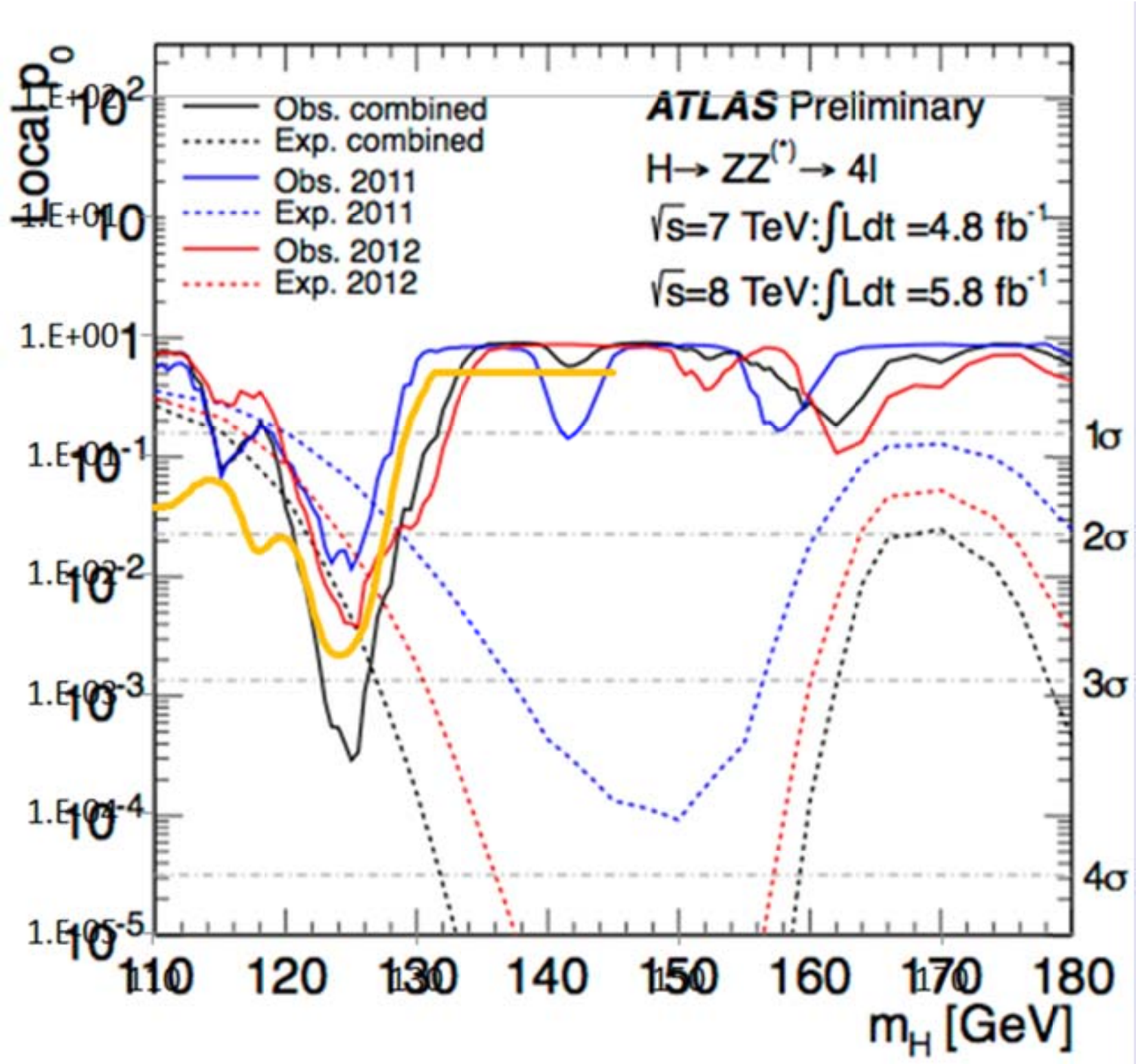

*Fig. 11- Local p-value of the four-lepton distribution from the present evaluation (yellow line) compared with the official ATLAS result (black line)*

*3.3 Joint local significance*

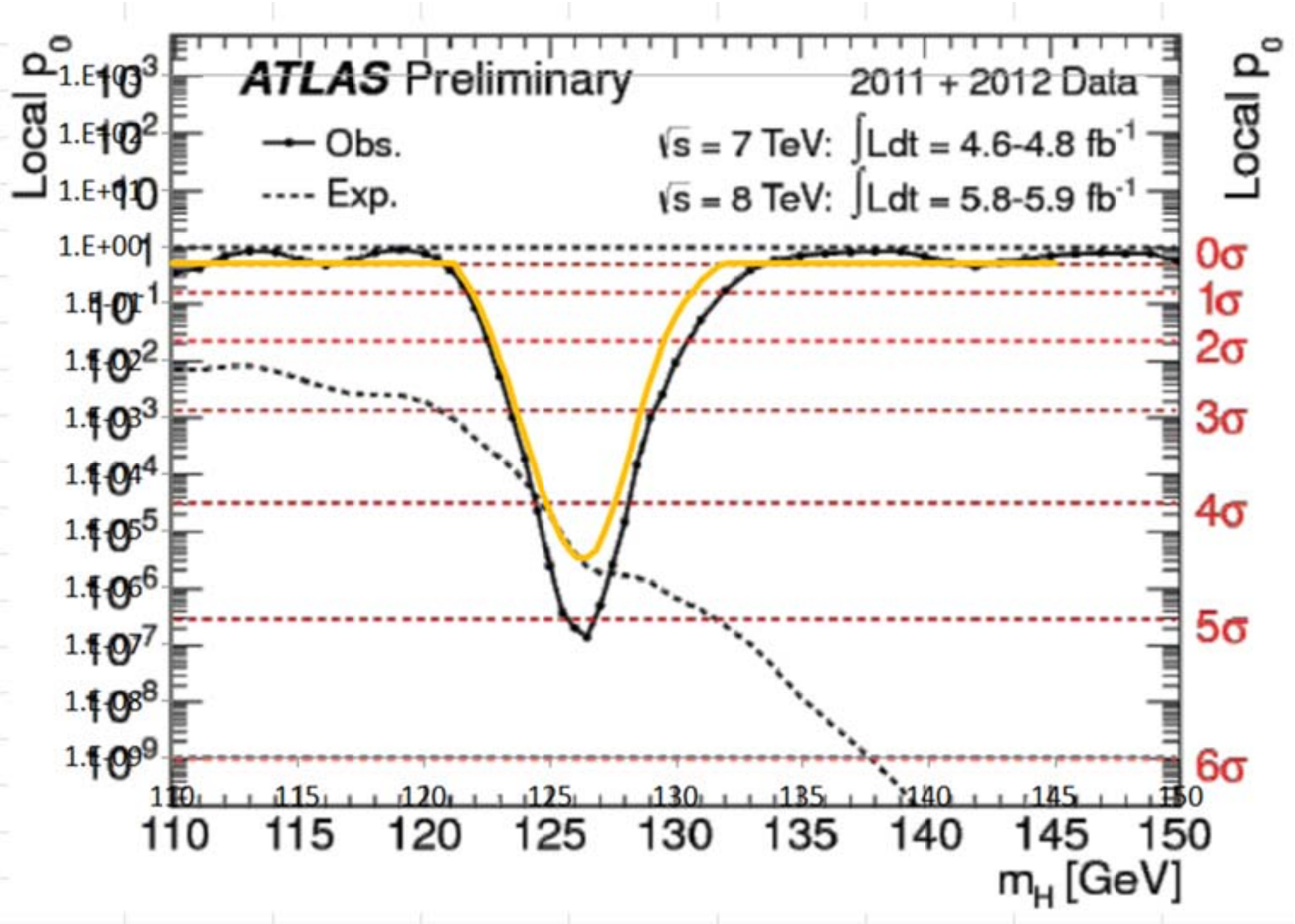

*Fig. 12 - Local p-value from the joint analysis of the four-lepton and gamma-gamma distributions through the present evaluation (yellow line) compared with the official ATLAS result (black line)*

The joint local p-value of the excess in the two considered channels is displayed in Fig. 12. The abscissa of the minimum obtained in the present exercise is 126.4 GeV, thus very well in agreement with the result of the Collaboration [11]. The minimum p-value corresponds to 4.5 σ, against the 5.1 official value of ATLAS. So, as for CMS, the result obtained here with a very simplified approach is somehow less, but not much less, significant than the official one; the higher significance of the official result is stemming from the enhanced sensitivity allowed by the categories analysis.

*3.4 LEE*

As for the CMS case, the present evaluation is completed with the determination of the Look Elsewhere Effect in the 110-146 GeV interval. The application of the same procedure reminded in § 2.4 originates a significance after the LEE of 3.7 σ, to be compared with the 4.2 σ value reported by the Collaboration in [11].

## 4. Conclusions

The (modest) purpose with which the writing of this note commenced was to try to understand which level of significance an external observer can gauge by himself/herself only exploiting the data released at the 4th July CERN seminars (and in the accompanying notes) by ATLAS and CMS.

Actually the result of this exercise is not bad: even though the 5 σ significance level cannot be attained with the simplified procedure employed, the indication of a strong effect, at the level of 4.5 -4.6 σ locally, is anyhow obtained for both datasets in the 125-126 GeV region. Moreover, the amount of de-rating of the significance due to the LEE, if one would consider the search over the entire low mass range, is consistent with de-rating obtained both by ATLAS and CMS.

This exercise confirms also that the indication of an excess in the two high resolution channels considered for this calculation, the gamma-gamma and four-lepton channels, is of very similar strength in both experiments.

**Acknowledgements**

I would like to thank for many enlightening discussions about the topic of this note, as well as other related topics, and for reading the manuscript, L. Mandelli, M. Fanti and A. Ianni.

## References

[1] CMS Collaboration, *Evidence for a new state decaying into two photons in the search for the standard model Higgs boson in pp collisions*, CMS PAS HIG-12-015 (2012)

[2] Joseph Incandela, *Status of The CMS SM Higgs Search,* CERN Seminar, 4th July 2012

[3] CMS Collaboration, *Evidence for a new state in the search for the standard model Higgs boson in the H →ZZ →4ℓ channel in pp collisions at √s = 7 and 8 TeV,* CMS PAS HIG-12-016 (2012)

[4] CMS Collaboration, *Search for the standard model Higgs boson in the decay channel H →ZZ →4ℓ in pp collisions at √s = 7 TeV,* Phys. Rev. Lett. 108, 111804 (2012)

[5] G. Ranucci, *The profile likelihood ratio and the look elsewhere effect in high energy physics,* Nuclear Instruments and Methods in Physics Research A, Volume 661, Issue 1, p. 77-85 (2012) , eprint: arXiv:1201.4604

[6] G.  Ranucci, *On the significance of the excesses in the ATLAS diphoton and four lepton decay channels,* arXiv:1201.6041

[7] CMS Collaboration*, Observation of a new boson with a mass near 125 GeV,* CMS PAS HIG-12-020 (2012)

[8] Fabiola Gianotti, *Status of Standard Model Higgs searches in ATLAS,* CERN Seminar, 4th July 2012

[9] ATLAS Collaboration, *Observation of an excess of events in the search for the Standard Model Higgs boson in the $\gamma\gamma$ channel with the ATLAS detector*, ATLAS-CONF-2012-091 (2012)

[10] ATLAS Collaboration, *Observation of an excess of events in the search for the Standard Model Higgs boson in the $H \to ZZ \to 4\ell$ channel with the ATLAS detector*, ATLAS-CONF-2012-092

[11] ATLAS Collaboration, *Observation of an Excess of Events in the Search for the Standard Model Higgs boson with the ATLAS detector at the LHC*, ATLAS-CONF-2012-093